%% file: main.tex
\documentclass[%
 reprint,
superscriptaddress,
nofootinbib,
 amsmath,amssymb,
]{revtex4-2}

\usepackage{graphicx}% Include figure files
\usepackage{dcolumn}% Align table columns on decimal point
\usepackage{bm}% bold math
\usepackage[colorlinks=true, allcolors=blue]{hyperref}
\usepackage{booktabs}
\usepackage{cancel}
\usepackage{braket}
\usepackage{multirow, tabularx}
\usepackage{amsmath, amsthm, amssymb}
\usepackage{comment}
\usepackage[toc,page]{appendix}
\usepackage[dvipsnames]{xcolor}
\usepackage{dsfont}
\usepackage{xcolor}
\usepackage{listings}
\usepackage{tabularray}
\usepackage[normalem]{ulem}
\usepackage{siunitx}
\UseTblrLibrary{booktabs}

\begin{document}

\preprint{}

\title{Postselection-loophole-free Bell test under strict spacetime constraints}
\textbf{}
% authors
\author{Kannan Vijayadharan}
% \email{kannan.vijayadharan@unipd.it}
\affiliation{Dipartimento di Ingegneria dell'Informazione, Universit\`a degli Studi di Padova, via Gradenigo 6B, IT-35131 Padova, Italy}

\author{Matías~Rubén~Bolaños}
\affiliation{Dipartimento di Ingegneria dell'Informazione, Universit\`a degli Studi di Padova, via Gradenigo 6B, IT-35131 Padova, Italy}

\author{Andrea Pompermaier}
\affiliation{Dipartimento di Ingegneria dell'Informazione, Universit\`a degli Studi di Padova, via Gradenigo 6B, IT-35131 Padova, Italy}

\author{Tommaso Bertapelle}
\affiliation{Dipartimento di Ingegneria dell'Informazione, Universit\`a degli Studi di Padova, via Gradenigo 6B, IT-35131 Padova, Italy}

\author{Francesco B. L. Santagiustina}
\affiliation{Dipartimento di Ingegneria dell'Informazione, Universit\`a degli Studi di Padova, via Gradenigo 6B, IT-35131 Padova, Italy}

\author{Costantino Agnesi}
% \email{costantino.agnesi@unipd.it}
\affiliation{Dipartimento di Ingegneria dell'Informazione, Universit\`a degli Studi di Padova, via Gradenigo 6B, IT-35131 Padova, Italy}
\affiliation{Padua Quantum Technologies Research Center, Universit\`a degli Studi di Padova, via Gradenigo 6B, IT-35131 Padova, Italy}

\author{Giuseppe Vallone}
\affiliation{Dipartimento di Ingegneria dell'Informazione, Universit\`a degli Studi di Padova, via Gradenigo 6B, IT-35131 Padova, Italy}
\affiliation{Padua Quantum Technologies Research Center, Universit\`a degli Studi di Padova, via Gradenigo 6B, IT-35131 Padova, Italy}

\author{Paolo Villoresi}
\affiliation{Dipartimento di Ingegneria dell'Informazione, Universit\`a degli Studi di Padova, via Gradenigo 6B, IT-35131 Padova, Italy}
\affiliation{Padua Quantum Technologies Research Center, Universit\`a degli Studi di Padova, via Gradenigo 6B, IT-35131 Padova, Italy}

\begin{abstract}
Entanglement gives rise to correlations between distant quantum systems that cannot be explained by local realistic theories. 
Bell inequality violations provide a direct way to reveal these correlations and certify nonlocality, especially when the relevant experimental loopholes are closed. 
Time-bin encoding, in which quantum information is encoded into well-defined temporal modes, is a commonly used platform for distributing photonic entanglement in optical fibers. 
Yet loophole-free Bell tests with time-bin entanglement have received comparatively little attention, owing in part to the postselection loophole introduced by conventional interferometric measurements.
Here, we demonstrate a fiber‑based platform for Bell tests with time-bin entanglement that simultaneously closes the locality, freedom-of-choice, and postselection loopholes. 
We observe a CHSH violation of $S=2.583 \pm 0.002$, exceeding the local‑realistic bound by over 265 standard deviations. 
Notably, this rigorous certification of nonlocality is achieved at a separation distance of $49.0\pm0.7$~m, substantially shorter than previous photonic Bell tests addressing comparable space-time constraints. 
Beyond its foundational significance, our results demonstrate time‑bin entanglement as a viable route towards practical device‑independent quantum communication and a future quantum internet.
\end{abstract}

\maketitle

\section{Introduction}

Device-independent (DI) protocols allow two or more parties to certify security or randomness without trusting the underlying hardware. 
This relies on the certification of genuine quantum correlations between the devices involved \cite{acinDeviceIndependent2007, BellNonlocalityWehner}. 
Quantum correlations can be certified by means of Bell tests, or Bell inequality violations, ruling out the possibility that the observed correlations could be explained by a local realistic theory. 
However, real-world implementations of Bell tests are limited by experimental constraints such as low detection efficiency, absence of true randomness, or possible subluminal communication between the parties when choosing the measurement settings. 
All of these experimental limitations, and more, introduce \emph{loopholes} in the test, where the statistics observed could in principle be reproduced by a classical local hidden variable model, thereby failing to rule out local realism.
This represents a severe issue for DI protocols, as the results of the test are not actually genuine.
Consequently, any DI protocol requires the Bell test to be performed in a loophole-free manner.

Intense research efforts have been made toward experimentally closing as many loopholes as possible, eventually achieving fully loophole-free scenarios. 
Notably, a loophole-free Bell test was demonstrated using an event-ready scheme on the electronic spin of NV centers with parties separated by 1.3 km \cite{hensenLoopholefree2015a}. 
Another loophole-free implementation was demonstrated using superconducting circuits separated by 30~m, requiring a highly advanced cryogenic microwave quantum link \cite{storzLoopholefree2023a}.
More recently, a device-independent implementation of quantum key distribution (QKD) was achieved with optical fiber spools as long as 100~km \cite{Lu2026}.
In the photonic domain, significant loophole-free tests have been demonstrated using the polarization degree-of-freedom with parties separated as little as 58~m \cite{giustinaSignificantLoopholeFree2015a, shalmStrong2015a}. 

When envisioning the future of deployed quantum networks, where multiple nodes separated by arbitrary distances must execute DI protocols, photonic systems are generally preferred due to lower transmission losses and higher resilience to decoherence compared to solid-state systems. 
The first and most intuitive degree of freedom used to encode entangled states is the photons' polarization, extensively demonstrated on deployed systems in free-space and fiber links. 
Limitations to its adoption arise in polarizing channels or, for long fiber links, from the emergence of polarization mode dispersion (PMD), which reduces the fidelity of entanglement, thus posing a significant challenge to fiber-based networks.
In addition, it is intrinsically bound to a bi-dimensional encoding.

An alternative that circumvents this limitation is the time-bin degree of freedom, which is largely immune to PMD, making it ideal for fiber implementations. 
Unlike polarization encoding, which is limited to two dimensions, time-bin encoded photons can be easily scaled to arbitrary higher dimensions. 
This scalability enables a new set of quantum protocols that can improve the noise resilience and potentially lower the critical detection efficiency requirements \cite{Huber2019Overcoming}.
Furthermore, time-bin encoding is particularly advantageous for heterogeneous quantum networks, facilitating interfaces with solid-state quantum nodes, such as matter-based quantum memories \cite{Iuliano2024, Knaut2024, Fischer2025}. 

Implementing time-bin receivers with a passive unbalanced interferometer (often called a Franson interferometer) introduces the post-selection loophole. 
Due to the probabilistic routing of the early and the late temporal modes into the long and short paths of the interferometer, the detection time histogram shows three peaks, where quantum interference occurs only in the central peak, representing 50\% of the total detected events.
To observe a Bell violation, a postselection procedure is required by the two parties, discarding all events except those where the photons arrive in the central peak, corresponding to a coincidence rate of 25\%.
Aerts et al. \cite{aertsTwoPhoton1999a} highlighted that this temporal filtering allows a local hidden-variable model to reproduce the quantum correlations through measurement-setting-dependent delays, effectively invalidating the Bell test. 
More recently, it was demonstrated that the postselection loophole renders time-bin and energy-time entanglement schemes vulnerable to quantum hacking attacks~\cite{Jogenfors2015}. 
This specific loophole has been studied and experimentally closed in~\cite{vedovatoPostselectionLoopholeFree2018b,santagiustinaExperimental2024a, Bernardi2026}. 
To the best of our knowledge, not much effort has been made to close the post-selection loophole in conjunction with other loopholes using time-bin entangled photons.

In this work, we report a Clauser–Horne–Shimony–Holt (CHSH) Bell test~\cite{PhysRevLett.23.880} experiment based on genuine time-bin entanglement in which the postselection loophole is closed, while simultaneously addressing the locality and freedom-of-choice loopholes in a photonic implementation.
With a spacelike separation of $\sim49$~m between measuring stations, referred to as Alice and Bob, our results demonstrate a loophole-aware Bell test at short separation distances that, to our knowledge, has not been achieved previously for any photonic Bell test.
Additionally, by integrating a Sagnac-modulator architecture into the standard Franson receiver, we enable fast optical switching as well as the basis choice without additional hardware. 
Beyond the reported violation, these high-speed Bell tests using time-bin entanglement under strict spacelike separation conditions represent a significant step towards scalable quantum networks and practical device independence.

\subsection*{Requirements for closing the addressed loopholes}
The CHSH Bell test consists of independent rounds, during which an entangled state is generated and sent to two independent parties: Alice and Bob. 
For each round of the test, Alice (Bob) chooses a measurement setting $x\in\{0,1\}$ ($y\in\{0,1\}$), which produces an outcome $a\in\{-1,+1\}$ ($b\in\{-1,+1\}$). Then, they estimate the set of conditional probabilities $P(a,b|x,y)$, from which the correlators are obtained as
\begin{equation}
    E_{xy}=\sum_{a,b=\pm1}ab\,P(a,b|x,y),
\end{equation}
which in turn are used to define the CHSH parameter
\begin{equation}
    S=E_{00}+E_{01}+E_{10}-E_{11}.
\end{equation}
If local realism holds for a loophole-free test, then $|S|\le2$, while quantum systems allow for $|S|\le2\sqrt{2}$, known as the Tsirelson bound. As such, the Bell test is deemed successful (i.e., local realism is ruled out) when $|S|>2$ with statistical significance.

To address the targeted loopholes, the conditions for their closure must be rigorously defined.
For the analysis of the loopholes that depend on the timing and distances in the Bell test, we define all events related to the test in a (2+1D) spacetime formalism\footnote{In our experimental layout, the source and receivers do not form a straight line, and consequently, the standard (1+1D) formalism is insufficient to fully describe the system.}, where each event $X$ is described by 2 dimensions in space and one in time, $X=(x,y,t)$, relative to the pair generation event $G=(0,0,0)$, which serves as the origin.
Then, we define $C_i=(x_C^i, y_C^i, t_C^i)$ as the event where party $i$'s measurement setting is decided, $S_i=(x_S^i, y_S^i, t_S^i)$ as the event where the measurement setting is actually applied to the photon received by party $i$, and $D_i=(x_D^i, y_D^i, t_D^i)$ as the event where that photon is detected. 

The \emph{locality loophole} arises when the spacetime configuration allows the choice of measurement setting or outcome at one measurement station to influence the outcome of the other. 
In this scenario, the observed correlations could, in principle, be explained by a local-realistic model where information about one party's setting or outcome reaches the other party before its measurements are completed. 
Closing this loophole therefore requires the relevant events at Alice and Bob to be spacelike separated, such that no subluminal signal can travel between them during the trial. 
We consider the locality loophole closed when satisfying the condition
\begin{equation}
    t_C^i + \frac{d_{A\leftrightarrow B}}{c} > t_D^j,\,\,\,\,\,i\ne j,
\end{equation}
for both $i,j=A,B$, with $d_{A\leftrightarrow B}$ the line-of-sight distance between the A and B parties and $c$ the speed of light in vacuum.
Moreover, if the test is performed in a nonlocal way, then the conditional probabilities must also follow 
\begin{equation}\label{eq:no-signaling}
    \begin{split}
        \sum_aP(a,b|x,y)&=\sum_aP(a,b|x',y),\\
        \sum_bP(a,b|x,y)&=\sum_bP(a,b|x,y'),
    \end{split}
\end{equation}
corresponding to no-signaling between the parties. 
This implies that the setting choice at Alice (Bob) cannot influence Bob's (Alice's) outcome.

When the choice of measurement settings for either party can be correlated with the hidden variables associated with the particle being measured, it is known as the \emph{freedom-of-choice loophole}. 
In this scenario, the assumption of statistical independence of the measurement settings from the source variables is no longer valid.
To close this loophole, the primary requirement is a source of genuine randomness for both parties to choose their measurement setting in each round of the Bell test. 
This is achieved by employing a quantum random number generator (QRNG) at each measurement station.
Under the assumption that any hidden variable $\lambda$, which can influence the measurement setting choice of either party, is created no earlier than the pair generation event, we consider the loophole closed when the setting choice event is spacelike separated from the source event, satisfying
\begin{equation}
    t_C^i < \frac{d_{S\leftrightarrow i}}{c},
    \label{eq:freedom_of_choice_dt_condition}
\end{equation}
where $d_{S\leftrightarrow i}$ is the line-of-sight distance between the entangled-pair source and the party $i$, where $t_C^i$ is defined relative to the pair generation event.

Besides the loopholes mentioned above, a fundamental loophole affects Bell tests where the states are encoded in temporal modes of light, such as time-bin or energy-time encoding \cite{aertsTwoPhoton1999a, jogenfors_2014}.
To measure these degrees of freedom, an interferometric measurement is required, where non-coherent detections are present, i.e., whenever the late (early) state takes the long (short) path of the unbalanced Mach-Zehnder interferometer (uMZI) at the receiver.
By discarding these events, only a subset that passes a postselection criterion is considered for the Bell test. This opens the \emph{postselection loophole}, where a local hidden variable $\lambda$ could, in principle, dictate which path of the interferometer is taken by the photon to maximize the result of the test.
There is more than one way to close this loophole, including special receiver interferometer topologies \cite{cabello_proposed_2009,santagiustinaExperimental2024a} and fast optical switches \cite{vedovatoPostselectionLoopholeFree2018b, Bernardi2026}. 
In particular, the fast optical switching solution proves to be more robust and easier to implement in a deployed network scenario.

Lastly, a loophole that we do not address in this work but is still worth mentioning is the \emph{detection loophole}, occurring when not all generated entangled states are effectively used for the Bell test. This happens in lossy scenarios where a significant portion of the generated photons is lost before detection. 
For maximally entangled states using standard CHSH operators, the critical overall detection efficiency required to close this loophole is $\eta_{\rm total} = 0.828$ \cite{Garg1987}, but can be lowered by moving to higher-dimensional systems \cite{Massar2002, Miklin2022}, using non-maximally entangled states \cite{Eberhard1993}, or using event-ready schemes \cite{Zeng2025}.
With our system, reaching these efficiency thresholds is not possible with state-of-the-art technology, and thus, we do not close the detection loophole. 
However, we stress that this is mainly a technological bottleneck and not a fundamental one, as low-loss optical components could improve this condition for our system.

\section{Methods}

In this section, we provide a detailed account of the experimental apparatus and validation methods underlying the results reported. We outline the performed Bell test, followed by details regarding the implementation of the entangled-photon source, the time-bin state analyzers, and the random number generation.
Additional details and supporting measurements are provided in the Supplementary Information.

\begin{figure*}[th!]
    \centering
    \includegraphics[width=\linewidth]{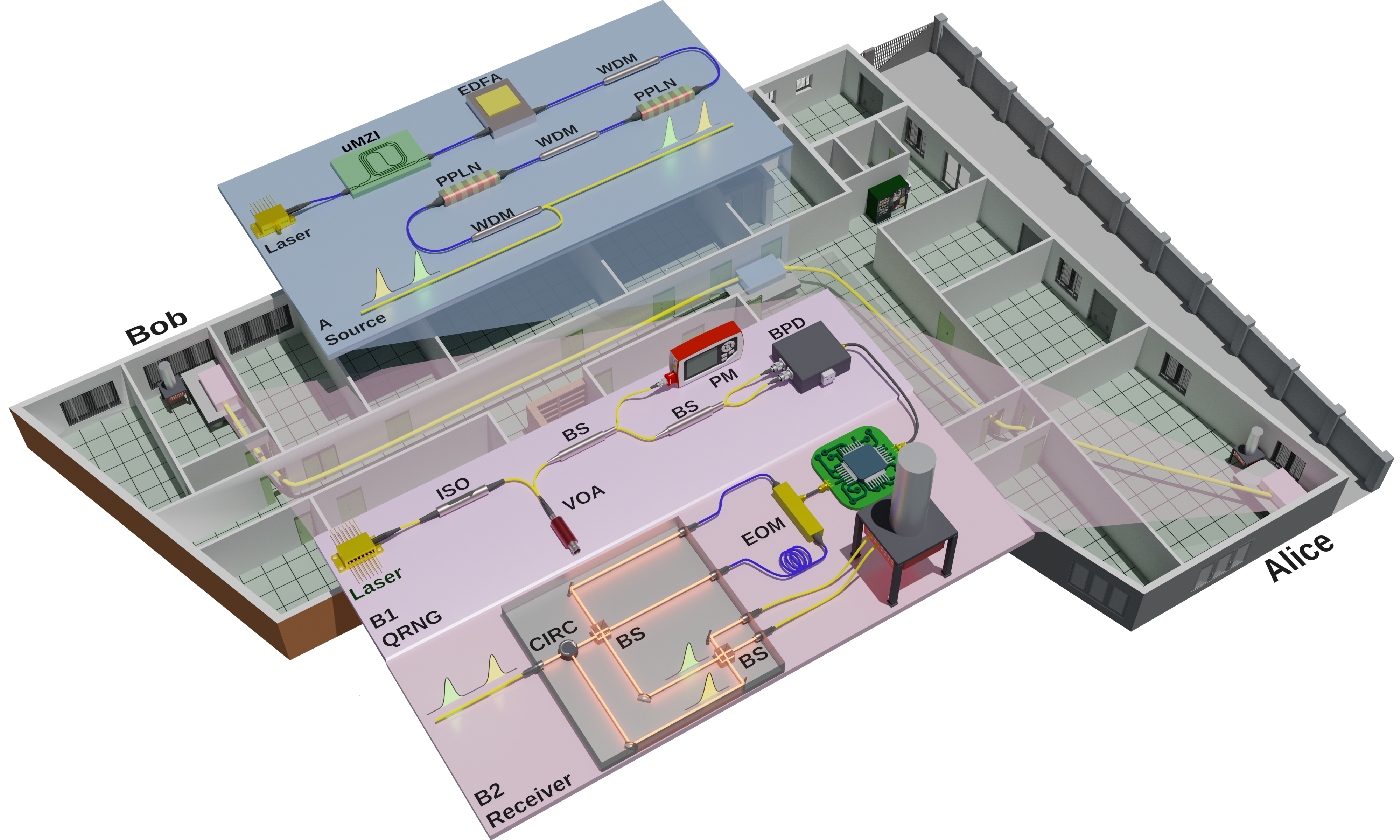}
    \caption{Experimental layout of the Bell test, qualitatively representing the positions of Alice and Bob (pink boxes) and the entanglement source (blue box) within the laboratory building, with the photon propagation paths from the source to each receiver shown by the single-mode fibers. (A) Entanglement source. (B) Each of the receiving stations, with (B1) representing the QRNG system and (B2) the time-bin receiver, implemented with a custom-made micro-optic assembly with fiber interfaces. uMZI: unbalanced Mach-Zehnder interferometer; EDFA: erbium-doped fiber amplifier; WDM: wavelength division multiplexing filter; PPLN: periodically-poled lithium niobate; ISO: isolator; VOA: variable optical attenuator; BS: beamsplitter; PM: power meter; BPD: balanced photodetector; CIRC: circulator; EOM: electro-optic modulator}.
    \label{fig:luxorMap}
\end{figure*}

The Bell test layout was designed to ensure spacelike separation between events at Alice and Bob ($d_{A\leftrightarrow B}=49.0\pm 0.7$~m), and to place the entanglement source symmetrically between the two parties (Fig.~\ref{fig:luxorMap}). 
A fiber-based photon-pair source employing a non-degenerate spontaneous parametric downconversion (SPDC) process was utilized to generate time-bin entanglement.
The source prepared the maximally entangled two-photon state $$\ket{\Phi_{+}} = \frac{1}{\sqrt{2}}\big(\ket{E}\ket{E}+\ket{L}\ket{L}\big)$$ at a repetition rate of $R_{\rm source}=125$~MHz, with the signal and idler photons directed to Alice and Bob respectively (see Supplementary Information for additional details).
After the source, both photons were propagated through single-mode fiber (SMF) to their corresponding party.
Both parties were equipped with a receiving station consisting of a continuous variable QRNG (CV-QRNG), an uMZI in which the input beamsplitter is replaced by a fast optical switch to close the postselection loophole, and a field-programmable gate array (FPGA) system that coordinates the measurements.
The QRNG device was designed to operate at 500~Mbit/s, generating 4 random bits per Bell trial (125~MHz).
Since only one bit was needed to choose between the two measurement settings for each party, the least significant bit was used to determine the setting, with the rest being discarded. More details regarding the QRNG characterization can be found in the Supplementary Information.

One way to close the postselection loophole is to use a fast optical switch to avoid non-interfering events, as previously realized by Vedovato et al. \cite{vedovatoPostselectionLoopholeFree2018b} using Mach-Zehnder modulators.
This architecture, however, requires active bias-point stabilization of the switch, adding complexity to long-duration operation. 
Here, we instead implement an intrinsically stable fast optical switch based on a Sagnac interferometer to perform both active switching and the measurement setting choice.
Additionally, by driving the electro-optical phase modulator nested inside the Sagnac switch according to a tailored delay and waveform profile, a phase between the early and late pulses can be applied, eliminating the requirement for an additional phase modulator to implement the measurement settings (see section \ref{sec:time-bin-receiver} for further details on the driving of the EOM), which is of particular importance when trying to minimize losses and delays.
After the optical switch, the photon continued through the uMZI, whose outputs were connected to a superconducting nanowire single-photon detector (SNSPD) system, whose outputs are tagged with 1 ps resolution and stored separately at each site. 

To perform the Bell test, we began with a calibration stage consisting of two steps.
First, we synchronized the switching at the receiver with the source by controlling the electrical signal driving the optical switch.
Then, we deterministically chose the measurement setting on each receiver, and the relative phase between the early and late states was set at the source to maximize the Bell parameter.
Following the calibration, the Bell test was initiated by triggering Alice and Bob to activate the QRNG and start measurements.
This resulted in a list of time-tagged detection events and QRNG-generated random bits for each party, which were subsequently post-processed to assign the measurement outcomes and evaluate the Bell test statistics. 

\subsection{Time-bin entanglement source}
The entangled photon pair source is divided into two main parts: a time-bin encoder and a pair-generation stage. 
In the first stage, a gain-switched discrete mode (DM) laser diode operating at $\lambda_{p} = 1560.61$~nm (ITU channel 21) generates a train of short optical pulses at a repetition rate of $R=125$~MHz. 
The time-bin superposition state is generated by passing the pulses through a planar lightwave circuit (PLC)-based uMZI. 
The phase difference between the two arms is controlled by the chip temperature.

After the encoding stage, the time-bin state is amplified by an Erbium-Doped Fiber Amplifier (EDFA), followed by wavelength-division multiplexing (WDM) filters centered on ITU-grid channel 21 to suppress out-of-band noise from the amplification. 
The amplified state is subsequently injected into the pair-generation stage, consisting of two \SI{30}{\milli\meter} type-0 periodically poled lithium niobate (PPLN) waveguides by HC Photonics. The first waveguide produces photons at $\lambda_{\rm SHG}=780.31$~nm through second harmonic generation (SHG), after which the residual pump light is removed using WDM filters. 
The second waveguide produces photon pairs through spontaneous parametric down-conversion (SPDC), and the signal and idler photons at wavelengths $\lambda_s= 1558.17$ nm (ITU-grid channel 24) and $\lambda_i=1563.05$ nm (ITU-grid channel 18) are spectrally filtered and distributed to the corresponding receiver through single-mode fibers.
The time-bin superposition state produced by the encoder is
\begin{equation}
    \ket{\phi}_p = \frac{\ket{E}_p +  \ket{L}_p}{\sqrt{2}} \,,
\end{equation}
where $\ket{E}$ and $\ket{L}$ denote the early and late time bins, respectively. In the pair-generation stage, the pump state $\ket{\phi}_p$ is converted, through the cascaded process, into the maximally entangled time-bin state
\begin{equation}
    \ket{\Phi_+} = \frac{1}{\sqrt{2}}\left( \ket{E}_s\ket{E}_i +  \ket{L}_s\ket{L}_i\right) .
\end{equation}
More details regarding the source characterization and its performance are provided in the Supplementary Information.

\subsection{Postselection-free time-bin receiver}\label{sec:time-bin-receiver}

The time-bin receiver at the Bell test measurement station is based on a customized fiber-coupled uMZI in a micro-optic assembly and lithium niobate EOM, providing a compact system compatible with telecom infrastructure.
For postselection-free measurements, the passive MZI is adapted into an active configuration by replacing the input beamsplitter with a fast optical switch synchronized to the source, allowing deterministic routing of the incoming time bins.
The optical path imbalance in the receiver interferometer matches the time-bin separation of $\Delta t=1$ ns, enabling two-photon interference.

The core component of each receiver is a Sagnac-based electro-optic switch that simultaneously enables fast switching of the time bins as well as phase encoding while maintaining intrinsically stable operation and a high extinction ratio. 
The switch is realized by placing a lithium niobate EOM asymmetrically in a Sagnac loop. 
Due to the asymmetry, the EOM interacts with the clockwise and counterclockwise propagating components at different times, allowing them to be modulated independently \cite{berra2025generalmodelmodulationstrategies}. 
The operation of the switch is given by the operator $\hat{U}_{\text {sw}}(\phi(t))$ acting on the input state such that
\begin{equation}
\hat{U}_{\text {sw}}(\phi(t))=e^{i \phi(t) / 2}\left(\begin{array}{cc}
\cos (\phi(t) / 2) & i \sin (\phi(t) / 2) \\
i \sin (\phi(t) / 2) & \cos (\phi(t) / 2)
\end{array}\right),
\end{equation}
where $\phi(t)$ is set such that the late (early) wave packet is deterministically routed to the short (long) arm of the uMZI. 

\begin{figure}
    \centering
    \includegraphics[width=1\linewidth]{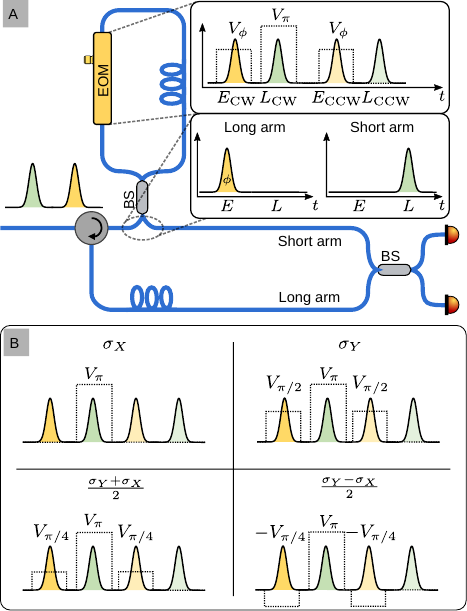}
    \caption{A) Scheme of the fast optical switch. A photon wavepacket is routed to the long arm of the interferometer when an arbitrary but equal phase is applied to both its clockwise (CW) and counterclockwise (CCW) components, and to the short arm when a $\pi$ phase shift is applied to only one component (here, the CW component). B) The four phase configurations implementing the four measurement operators required for the CHSH test, according to Eq.~\eqref{eq:switch-operator}.}
    \label{fig:switch-scheme}
\end{figure}

By routing the early wave packet ($\ket{E}$) to the long arm and the late wave packet ($\ket{L}$) to the short arm, the lateral peaks in the detection histogram are suppressed, resulting in a postselection‑free measurement, as shown in Fig.~\ref{fig:switch-scheme}A. The resulting histogram shows a dominant central peak (from the indistinguishable events $\ket{E}\ket{L}+\ket{L}\ket{E}$) with well-suppressed side peaks (Fig. \ref{fig:switch-histogram}), confirming the negligible impact of the side-peak amplitudes on the Bell test. 

A major advantage of the Sagnac switch is that, in addition to the switching action, it can implement the relative phase encoding of the same time-bin qubit being actively routed. The asymmetric configuration of the Sagnac switch is exploited in the RF drive waveform to enable this dual use, removing the need for a second EOM. 
This applied relative phase changes the measurement basis implemented by the receiver to project onto the eigenstates of the operator
\begin{equation}\label{eq:switch-operator}
    M(\phi)=\cos(\phi)\sigma_X-\sin(\phi)\sigma_Y.
\end{equation}
Thus, Alice randomly selects one of the two basis settings $\sigma_X$ and $\sigma_Y$, corresponding to $\phi_0^A=0$ and $\phi_1^A=\pi/2$ respectively, and Bob between the two basis settings $\frac{\sigma_X+\sigma_Y}{\sqrt{2}}$ and $\frac{\sigma_Y-\sigma_X}{\sqrt{2}}$, corresponding to $\phi_0^B=\pi/4$ and $\phi_1^B=-\pi/4$ respectively.
Experimentally, the fast basis selection is made trial-by-trial by a QRNG, and the corresponding electrical signal is generated by a high sampling rate digital-to-analog converter and amplified to the operating voltage levels of the EOM (Fig.~\ref{fig:switch-scheme}B).

\subsection{Quantum Random Number Generator}
The QRNGs adopted relied on continuous variable (CV) homodyne detection of vacuum field fluctuations~\cite{Gabriel2010}.
Indeed, by using a few components, such a scheme can reach high generation rates~\cite{Pan2021, Bruynsteen2026} and low latencies with the appropriate electronics and post-processing.
A schematic representation of the QRNG design is reported in Fig.~\ref{fig:luxorMap}: a 50:50 BS mixes the light of a strong laser with the vacuum state, and the resulting outputs are connected to a balanced photodetector (BPD) extracting one quadrature of the vacuum field.
Before entering the BS, the laser light passes through an optical isolator (ISO) to prevent backscattering onto the laser and a variable optical attenuator (VOA) to minimize saturation and non-linear effects in the BPD.
The BPD RF signal is then amplified to properly exploit the subsequent ADC quantization range and digitized by a \SI{4}{\giga Sa /\second} 12-bit resolution ADC integrated within an Ultrascale+ MPSoC development board.
Finally, the FPGA accumulates a block of 32 12-bit samples to be used as input to a 2-universal hashing function random extractor~\cite{Tomamichel2011}, implemented as a Toeplitz matrix-vector multiplication modulo two~\cite{Krawczyk1995}, from which 4 random bits are extracted.

The implemented procedure ensures that the extracted numbers are independent and follow a distribution $\varepsilon$-close to uniform, regardless of the input statistics.
Its value is determined by the quantum conditional min-entropy per homodyne measurement, the number of measurements (block) to hash, and the size of the resulting compressed block.
It is worth noting that the $\varepsilon$-closeness parameter applies to a single hashed block, and if the same Toeplitz matrix is used for N blocks, the overall closeness decreases linearly with N, meaning $\varepsilon^\prime = N\varepsilon$.
Given the matrix-vector multiplication structure of Toeplitz extraction, it is ideal to be implemented on FPGA architectures, as all operations can be mapped to bitwise operations, which are native and optimized for the technology.

Using the homodyne QRNG scheme, we built two devices that meet the required \SI{125}{\mega b/s} generation rate, with a temporal correlation window between homodyne measurements, due to the low-pass behavior of the receiver’s analog electronics, of less than \SI{1.48}{\nano\second}.
Combined with the device latencies due to readout and post-processing, this keeps us within the timing constraint needed to ensure that the freedom-of-choice loophole is closed.
Moreover, with a hashing matrix of dimensions $384$-by-$4$ and using only the least significant bit of the hashed data, the single-shot $\varepsilon$ parameters are approximately $10^{-49}$ and $10^{-48}$, leading to an overall $\varepsilon^\prime$ closeness parameter of $\sim 10^{-37}$ and $\sim 10^{-36}$, given the 1650~s duration of the Bell test.

\subsection{Data processing}

During the experiment, each measurement station independently records detection events as $(t_n^i,c_n^i)$ pairs, where $t_n^i$ is the detection time in picoseconds for round $n$ at party $i$, and $c_n^i$ encodes the detector channel for the event.
In parallel, the full QRNG bit stream is generated at 125 Mb/s and stored in a binary file within the FPGA to minimize overhead.

Although the time-tagging devices share a synchronized clock, an offset arising from the device startup times required correction. 
This was addressed through cross-correlation analysis over detections registered before the beginning of the test, followed by manual optimization to center the main coincidence peak at $t=0$.

To evaluate the Bell test in a postselection loophole-free manner, coincidences are defined by the source pulse index rather than a time difference threshold.
Every aligned tag is assigned a pulse index $n$ according to
\begin{equation}
    n(t_n^i) \;=\; \mathrm{round}\!\left(\frac{t_n^i}{T_s}\right),
\end{equation}
where $T_s = 8$~ns is the source repetition period.
A coincidence is registered whenever tags from Alice and Bob share the same pulse index $n$.
All coincidences are accounted for in the Bell test, and the discarded events are the ones where only one party recorded a detection; this is acceptable as the test was performed under the fair-sampling assumption.

At each measurement station, the QRNG generates random bits at 125 Mb/s, with the full sequence stored on disk. 
Additionally, every $100$th bit is routed to a dedicated time-tagger channel to establish a timing reference between the bit stream and the detection events. 
From the time-tagged events, a subset of the generated random bits is reconstructed and then cross-correlated with the full bit sequence stored on disk to determine the global QRNG offset for Alice and Bob independently. 

For each coincidence at pulse index $n$, the measurement settings $x_n, y_n \in \{0,1\}$ chosen by Alice and Bob are read from the aligned QRNG records at indices $n - \Delta_A$ and $n - \Delta_B$, respectively. 
The constants $\Delta_{A,B}$ represent the fixed FPGA-to-detector latency and were determined, together with the QRNG offsets, by maximizing $|S|$ on an initial calibration window. 
Once fixed, they remain constant for the duration of the experiment, and the data from the calibration window is discarded from the test.

With both time-tags and QRNG bit streams aligned, every coincidence observed after the calibration window is used to perform the CHSH test.
For each setting pair $(x,y)$, we count the number of coincidences $N_{ab}(x,y)$ for the four outcome combinations $a,b=\pm1$, and estimate the value of the correlators
\begin{equation}
    E_{xy} \;=\; \frac{\sum_{a,b=\pm1} ab \cdot N_{ab}(x,y)}{\sum_{a,b=\pm1} N_{ab}(x,y)},
    % \frac{N_{++} - N_{-+} - N_{+-} + N_{--}}{N_{++} + N_{-+} + N_{+-} + N_{--}},
\end{equation}
shown in the inset of Fig.~\ref{fig:correaltors}, with standard deviation given by
\begin{equation}
    \sigma  = \sqrt{\frac{1 - E_{xy}^2}{\sum_{a,b=\pm1}N_{ab}(x,y)}}\, .
\end{equation}

As previously mentioned, the set of no-signaling conditions constitutes a necessary condition for closure of the locality loophole.
However, since non-coincident events are discarded for this test, the marginal conditional coincidence probabilities for one party are not necessarily independent of the other; the estimated probabilities are conditioned on a recorded coincidence and thus depend on asymmetries such as detector efficiency mismatch. 
Indeed, when estimating probabilities in this manner, we observed statistically significant differences between marginal probabilities, with the worst-case difference being 
\begin{equation}
    \Delta_{a=+1}=|P(+1|x,0) - P(+1|x,1)|=0.006\pm0.001,
\end{equation}
almost six standard deviations above zero. 
To ensure that the no-signaling conditions held, the value reported in the main text is obtained by estimating the conditional probabilities prior to filtering for coincidences.

\section{Results}

\begin{figure}[ht!]
    \centering
    \includegraphics{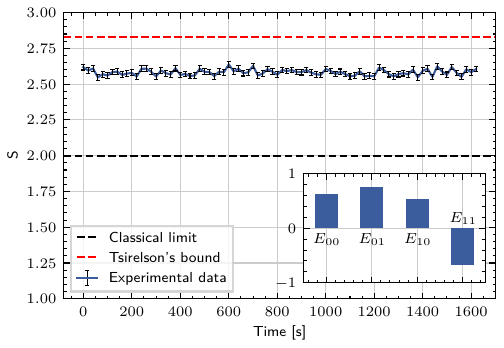}
    \caption{Measured $S$ parameter during $1650$ seconds of acquisition. Each point represents the $S$ parameter determined by 20 seconds of integration. Inset: CHSH correlators of the Bell test for the entire measurement.}
    \label{fig:correaltors}
\end{figure}

Multiple independent Bell tests were performed to assess robustness and reproducibility of the system. 
All measurements yielded consistent results, with variations below 5\%; here we report the data from a representative measurement lasting for about 1650~s.
The Bell test was performed by accumulating around $1.9 \times 10^{6}$~coincidence events while closing the loopholes addressed in the previous sections.
The correlators for each measurement setting combination were estimated over the integration period, from which a Bell parameter $S=2.583\pm 0.002$ was obtained (Fig.~\ref{fig:correaltors}), achieving a value above the local realism limit by over 265 standard deviations of statistical significance.
We note, however, that this result relies on discarding `no-click' events at both measurement stations, implying that this result holds under the fair-sampling assumption.
The measured conditional coincidence probabilities $P(a,b|x,y, \text{coinc.})$ are consistent with a measurement model in the equatorial plane of the Bloch sphere. More details are provided in section \ref{sec:chsh-model}.

Finally, the set of no-signaling conditions shown in Eq.~\eqref{eq:no-signaling} constitutes a necessary requirement for claiming closure of the locality loophole.
The conditional probabilities were calculated from the full dataset for each party prior to filtering for coincidences.
As expected, this method resulted in adherence to these conditions, with the worst-case difference being
$$\Delta_{a=+1}=0.0002\pm0.0002,$$
indicating a decrease in the absolute value by an order of magnitude and confirming that the result contains zero within only one standard deviation.

\subsection{Certifying loophole closure}

\begin{figure*}
    \centering
    \includegraphics{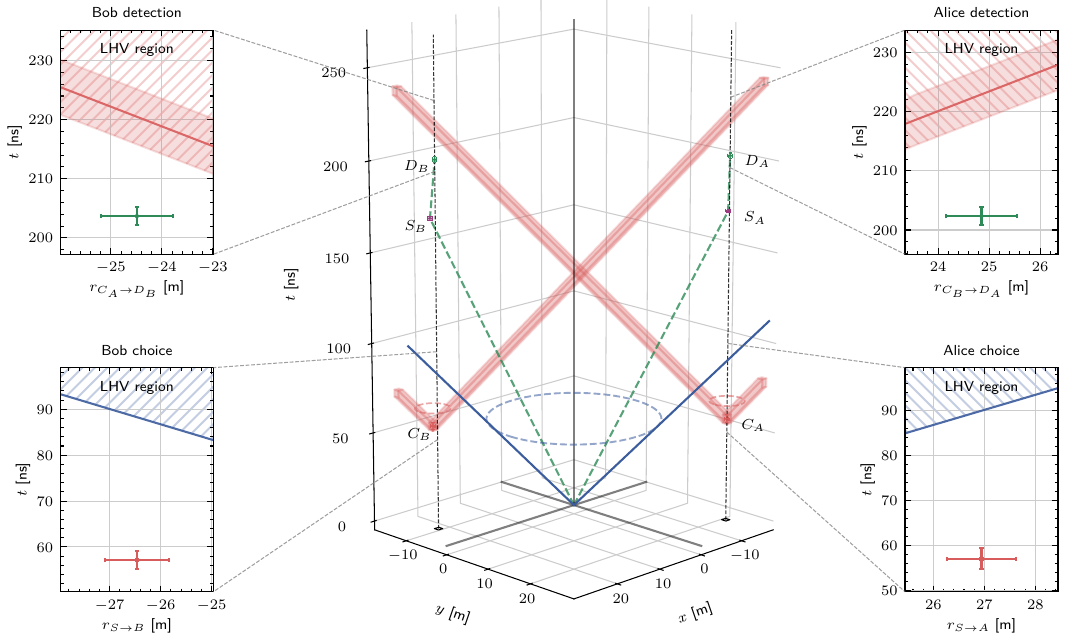}
    \caption{(2+1D) space-time diagram of the Bell test events relative to the generation of an entangled pair. The spatial and temporal coordinates for all represented events are written explicitly in the Supplementary Information. The future light-cone of the pair generation event is represented by solid blue lines, while the light-cone of the measurement setting choice is shown by extruding the entire event span in the direction of the other party's detection event.
    The dashed green line represents the actual photon propagation from the pair generation to its detection.
    The zoomed-in insets are represented as (1+1D) events by projecting the 2 spatial dimensions to the line joining both positions being shown.}
    \label{fig:minkowski}
\end{figure*}

First, the propagation time of the photon from the source to the EOM of each receiver is characterized by fiber propagation time and was measured to be $t_{\rm prop}^{S\to A}=172\pm1$~ns and $t_{\rm prop}^{S\to B}=172\pm1$~ns, from which the temporal coordinate of the events $S_i$ for the two parties were obtained.
Then, taking into account the electrical and optical latencies of the QRNG and FPGA systems for party $i$, $t_{\rm setting}^i=t_S^i-t_C^i$, we can define the events $C_i$ for both parties, corresponding to $t_{\rm setting}^A=115\pm 1$~ns and $t_{\rm setting}^B=115\pm 1$~ns.
Then, the photon propagation delay from the measurement setting, performed by the EOM, to the detection event, $t_{\rm detection}^i=t_D^i-t_S^i$ was measured to be $t^A_{\rm detection}=30.4\pm 0.5$~ns and $t^B_{\rm detection}=32.1\pm 0.5$~ns.
All relevant spacetime events are shown in the Minkowski diagram in Fig.~\ref{fig:minkowski}, with timing and position uncertainties evaluated under conservative worst-case assumptions.
The way in which we estimated all of these timings and positions, and their uncertainties, is further explained in the Supplementary Information.
To the best of our knowledge, this represents the shortest distance achieved in a photonic Bell test closing the locality and freedom-of-choice loopholes, as summarized in Table~\ref{tab:distacnesbelltest}.

\begin{table}[t]
    \centering
    \begin{tabular}{lcc}
        \toprule
        \textbf{Year} & \textbf{Platform} & $\boldsymbol{d_{A\leftrightarrow B}}$ \textbf{[m]} \\
        \midrule
        2010 & Photonic (polarization)\cite{scheidlViolation2010} & 143,600 \\
        2015 & NV center \cite{hensenLoopholefree2015a} & 1,280 \\
        2015 & Photonic (polarization) \cite{shalmStrong2015a} & 184.9 \\
        2015 & Photonic (polarization) \cite{giustinaSignificantLoopholeFree2015a} & 58 \\
        2018 & Photonic (polarization) \cite{liTest2018} & 184.9 \\
        2018 & Photonic (polarization) \cite{rauchCosmic2018} & 1028 \\
        2023 & Superconducting qubit \cite{storzLoopholefree2023a} & 32.824 \\
        2024 & Photonic (polarization) \cite{ZhaoLoophole-Free2024} & 183* \\
        \textbf{2026} & \textbf{Photonic (time-bin), this work} & \textbf{49} \\
        \bottomrule
    \end{tabular}
    \caption{Overview of relevant loophole-aware Bell test experiments across various platforms and encoding methods. *Distance estimated using the reported values of $d_{S\leftrightarrow A}$ and $d_{S\leftrightarrow B}$.}
    \label{tab:distacnesbelltest}
\end{table}

\begin{figure}[!ht]
    \centering 
    \includegraphics{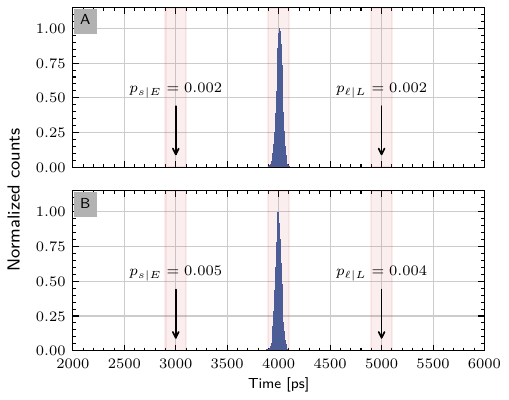}
    \caption{Histogram of detection events with the switch enabled for both (A) Alice and (B) Bob. The plots illustrate the estimated probability of `incorrect' routing, denoted as $p_{s|E}$ ($p_{\ell|L}$), representing the probability that an early (late) photon takes the short (long) arm of the receiver interferometer.}
    \label{fig:switch-histogram}
\end{figure}

To address the postselection loophole, the performance of the self-compensating Sagnac-based optical switch was evaluated to ensure that analyzing all detected coincidence events (i.e., without temporally filtering or discarding any) has negligible impact on the visibility. 
From this, we observed a negligible probability of detecting photons in the side peak slots for both receivers, as shown in Fig.~\ref{fig:switch-histogram}.
This was limited by the detectors' dark count rate and therefore provides a conservative upper bound on the actual error probability. 

\subsection{Effective model for the CHSH outcomes}\label{sec:chsh-model}

\begin{figure*}[t!]
    \centering
    \includegraphics{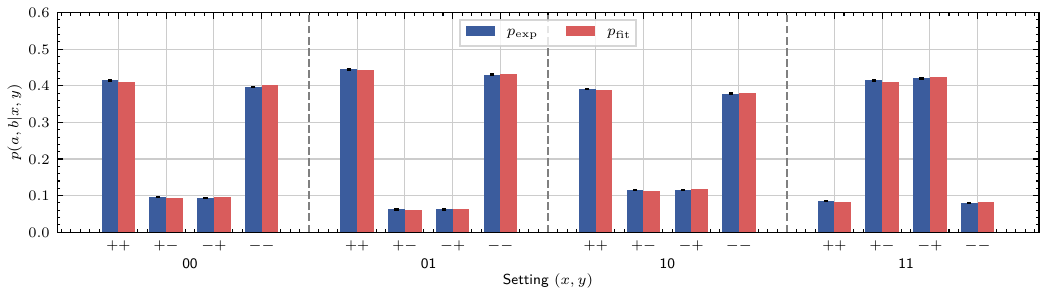}
    \caption{Measured (blue bars) and fitted (red bars) outcome probabilities. Experimental coincidence-conditioned probabilities ($P(a,b|x,y,\text{coinc.})$) are shown for all outcome pairs $(a,b\in{+1,-1})$ and measurement settings $(x,y\in{0,1})$, together with the probabilities predicted by the fitted theoretical model.}
    \label{fig:probs-fit}
\end{figure*}

We model the measured coincidence statistics using an effective two-photon interference model that includes finite visibility, errors in the applied measurement phases, and detector efficiency mismatch.
For outcomes $a,b\in\{+1,-1\}$ and basis choices $x,y\in\{0,1\}$, the probabilities are given by
\begin{equation}
    P(a,b|x,y, \text{coinc.}) \propto {\eta^A_{a}\eta^B_{b}[1+ab\mathcal{V}\cos(\Theta_{xy})]},
\end{equation}
where
\begin{equation}
    \Theta_{xy}=(\phi_{A,x} +x\delta_A) + (\phi_{B,y} +y\delta_B) - \phi_0
\end{equation}
is the joint state analyzer phase. Here, $\phi_{A,x}$ and $\phi_{B,y}$ denote the measurement basis phases chosen by Alice and Bob; $\mathcal{V}$ represents the effective visibility accounting for interferometer delay mismatch, multi-pair emission from the SPDC source, and phase drift during the measurement; the relative phase offsets $\delta_{A}$ and $\delta_{B}$ account for errors in the applied local measurement settings, while $\phi_0$ denotes the global phase offset of the entangled state relative to the receivers.
The relative output-dependent efficiencies were calculated directly from the recorded detection events to be $\eta^A_{+1}=0.995\,\eta^A_{-1}$ and $\eta^B_{-1}=0.973\,\eta^B_{+1}$. 
The remaining model parameters were obtained by numerical optimization as $\mathcal{V}=0.92$, $\delta_A=0.2$, $\delta_B=-0.13$, $\phi_0=-0.04$.
The model captures the dominant probability structure of the data (as shown in Fig. \ref{fig:probs-fit}), with $R^2=0.99969$ for the measured versus fitted conditional probabilities.

By setting the visibility to unity ($\mathcal{V}=1$) while keeping all other optimized parameters constant, the expected $S$ parameter increases to $S=2.809$, which corresponds to \SI{99.3}{\percent} of the theoretical maximum established by the Tsirelson bound. 
This indicates finite visibility as the dominant contributor to the gap between our experimental result and the theoretical maximum, limited by the non-adjustable temporal delay mismatch between the source and receiver uMZI.

\section{Conclusions and outlook}

Although a significant Bell violation of over 265 standard deviations was observed, the current implementation remains limited by several technological constraints.
The primary bottleneck is the finite bandwidth and dynamics of the electro-optic switching architecture, as performing both switching and basis choice in the same EOM affects signal integrity and restricts the observed visibility. 
Moreover, the high insertion losses of the commercial traveling-wave EOMs contributed to a majority of the system losses. By transitioning to suitable low-loss \cite{Pereira2022Electrooptic} and all-optical switching architectures \cite{Kupchak2019Terahertz, Couture2025Terahertz}, the system performance could be scaled and losses minimized to potentially also close the detection loophole.

Secondly, while the Sagnac-based optical switches enabled active routing of the time bins without requiring bias drift compensation, the configuration made it challenging to stabilize the uMZI itself by propagating a pilot signal backwards. 
This issue was largely mitigated by employing an athermal micro-optic assembly \cite{PhysRevApplied.7.044010} for the receiver interferometer, which provided passive thermal stability over long measurement durations. 
Alternatively, the receiver could leverage thin-film lithium niobate integrated photonic platforms \cite{Bernardi2026} for high-bandwidth switching and phase stability.
Additionally, the repetition rate was limited by the experimental architecture performing active basis choice and switching with the same EOM.

Finally, to further tighten the spacelike separation conditions, the propagation channel can itself be optimized. 
By replacing standard single-mode fibers with anti-resonant hollow-core fibers \cite{Petrovich2025, Antesberger:24}, the index of refraction can be significantly reduced, thereby lowering the photon propagation time. 

Addressing the limitations discussed in this section will be essential for scaling this architecture towards the high-rate, fiber-based networks required for practical device-independent quantum information processing. 
The time-bin receiver architecture introduced here allows for postselection-free measurements and high-speed active basis choice scalable to the gigahertz regime and to high-dimensional Bell tests. 
Additionally, the experimental platform exploits a low-latency QRNG, which allowed for independent measurement setting choice at 125~MHz. 
Combined, this has enabled the closure of locality, freedom-of-choice, and postselection loopholes for time-bin entangled photons with an Alice-Bob separation of less than 50 m. 

Our results make loophole-free Bell tests based on time-bin entanglement more accessible and show that device-independent quantum communication protocols, the gold standard for this field, are within reach. 
Indeed, we show here a large violation of a Bell inequality with a significance of 265 standard deviations while closing the locality, freedom-of-choice, and postselection loopholes. 
This approach paves the way for fundamental studies using time-bin entangled states to investigate the nature of physical reality, as well as for practical applications of entanglement distribution in quantum networks.

\bibliography{ref}% Produces the bibliography via BibTeX.

\subsection*{Author contributions}
K.V. conceived the idea for the experiment. 
K.V., M.R.B. and C.A. designed the system and the experiments. 
M.R.B. designed the control electronics. 
A.P. designed the photon pair source.
T.B. designed the QRNG system.
K.V., M.R.B., A.P., T.B., and C.A. performed the measurements and system characterization, and data analysis. 
C.A., G.V., and P.V. supervised the experiments. 
All authors discussed the results and wrote the manuscript.

\begin{acknowledgments}
K.V. acknowledges support from the European Union’s Horizon Europe Framework Programme under the Marie Sklodowska Curie Grant No. 956071, Project AppQInfo, and from the John Templeton Foundation by grant No. 63132, as recipient of an Enrico Fermi Fellowship awarded through the Center for SpaceTime and Quantum. 
M.R.B. acknowledges support from the European Union’s Horizon Europe Framework Programme under the Marie Sklodowska Curie Grant No. 101072637, Project Quantum-Safe Internet (QSI).

The opinions expressed in this publication are those of the author(s) and do not necessarily reflect the views of the respective funding body.

The authors would like to thank Alain J. Corso, Giulio Favaro, and Marta Padovani for allowing us to use their laboratory to set up a measurement station for the Bell test. 
\end{acknowledgments}

\input{supplementary}

\end{document}

%% file: supplementary.tex
\clearpage
\onecolumngrid
\setcounter{subsection}{0}
\section*{Supplementary Information}

\subsection{\label{sec:performance-characterization}Source characterization}

Experimental results indicated a reduction in the expected CHSH $S$ parameter, attributed primarily to the non-idealities of the entangled pair source. This section describes the characterization of the source in the actual Bell test configuration to quantify its impact.

\begin{figure*}[!ht]
    \centering
    \includegraphics{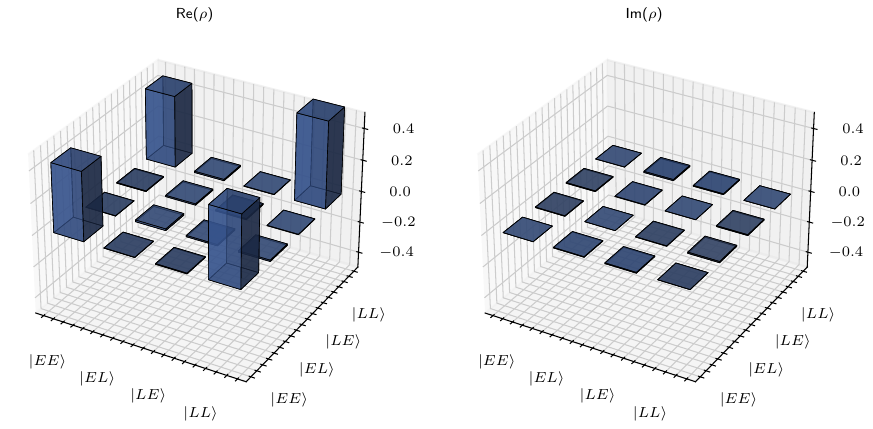}
    \caption{Reconstructed density matrix of the state by quantum state tomography.}
    \label{fig:densityMatrix}
\end{figure*}

The SPDC source generates correlated photon pairs with a heralding efficiency of 46.7\% for the signal (C24) and 50.5\% for the idler (C18).
The maximally entangled time-bin state created by the source was characterized through quantum state tomography. 
To reconstruct the full two-qubit density matrix, measurements were performed in the three mutually unbiased Pauli bases, $\sigma_X$, $\sigma_Y$ and $\sigma_Z$, for both Alice and Bob. The time-bin receivers with the fast optical switch allowed measurements in all three bases. 
The $\sigma_Z$ basis corresponds to direct time-of-arrival measurements, distinguishing the early and late time bins. This was realized by keeping the optical switch off, consequently routing both bins to the same arm of the interferometer. 
Instead, the $\sigma_X$ and $\sigma_Y$ measurements were implemented in the superposition basis with the switch on and setting the receiver phases to the corresponding equatorial measurement bases. 
Coincidence counts were acquired for the nine basis combinations by applying the measurement sequence and were used to reconstruct the two-qubit density matrix by maximum-likelihood estimation constrained to be positive semidefinite and unit trace (Fig. \ref{fig:densityMatrix}).
With this procedure, a state fidelity of $F =\bra{\Phi} \rho \ket{\Phi} = 0.9478 \pm 0.0007$ was estimated with respect to the state $\ket{\Phi} = (\ket{EE} + e^{i\phi}\ket{LL})/\sqrt{2}$. 
The infidelity is mainly attributed to reduced interferometric visibility arising from both an imbalance between the $\ket{E}$ and $\ket{L}$ time-bin populations and a mismatch between the source time-bin separation, with additional contributions from $\sigma_Y$ basis phase calibration errors and accidental coincidences from the SPDC source. The reconstructed state is nevertheless consistent with the visibility obtained in the Bell test measurements.

\subsection{QRNG performance characterization}\label{appendix:qrng_perf_characterization}

In this appendix, we will go over the methods used to determine the QRNG parameters: the extractable randomness per measurement, the security parameter $\varepsilon$ that measures how close the extracted randomness is to a uniform distribution, and the temporal correlation window of the homodyne receiver's electronics (analog backend), which indicates how many past events influence the homodyne outcome at time $n$.
It is worth mentioning that the first parameter directly affects the generation rate of the device, the second the deviation of the post-processed distribution from the ideal uniform one, while the last the amount of time that we need to push the QRNG light cone back in time due to the device's analog back-end affecting the freedom-of-choice loophole.

Regarding the randomness that can be extracted per measurement, the device employs a fully trusted homodyne protocol that measures the vacuum fluctuations.
The lower bound on the amount of true randomness that can be extracted per homodyne measurement is given by
\begin{equation}
    H_{\rm min} (X \vert E) = -\log_2 p_{\rm max} \Delta x \, ,
\end{equation}
where $p_{\rm max}$ is the maximum of the Gaussian distribution due to the vacuum field, and $\Delta x$ is the homodyne receiver resolution (which is ultimately set by the FPGA's ADC).
Nevertheless, directly using $p_{\rm max}$ obtained from the raw homodyne data is not adequate.
In fact, our goal is to extract the true randomness generated solely by the quantum process.
Yet this quantum contribution is mixed with extra noise originating from the receiver’s electronics.
Since such a noise is considered classical, its contribution to the min-entropy must be removed.
Additionally, computing the power spectral density (PSD) of the homodyne measurement shows that it is not flat.
This implies the presence of correlations within the measurements that, if ignored, may cause the estimated value of $H_{\rm min}(X \vert E)$ to deviate statistically from its actual value.
To tackle the aforementioned issues, we proceed as follows.
We assumed that the homodyne signal is of the form
\begin{equation}
    y_n = x_n + e_n ,
\end{equation}
where $x_n$ and $e_n$ are two mutually independent stationary temporally correlated random processes associated, respectively, with the quantum signal and the background electronic noise.
Since both $x_n$ and $e_n$ are normally distributed, $y_n$ is also Gaussian. 
Therefore, by having experimental access to $y_n$ (the raw homodyne measurements) and $e_n$ (the electronic background noise acquired when the homodyne LO was turned off), we can estimate the variance of the distribution $x_n$ conditioned over the past measurements (which is only associated with the quantum process measured at time $n$) as
\begin{equation}
    \sigma_{x\vert {\rm past}}^2 = \sigma_{y\vert {\rm past}}^2 - \sigma_{e\vert {\rm past}}^2\, ,
\end{equation}
where $\sigma_{y\vert {\rm past}}^2$ and $\sigma_{e\vert {\rm past}}^2$ are the conditional variances of $y_n$ and $e_n$ over past outcomes.
As we assumed the random processes to be stationary, $\sigma_{y\vert {\rm past}}^2$ and $\sigma_{e\vert {\rm past}}^2$ can be derived directly from the PSDs of $y_n$ and $e_n$ $W_{y,e}(\nu)$ as~\cite{alma990004119180206046}
\begin{equation}
    \sigma^2_{y,e | {\rm past}} = \exp{\int_{-1/2}^{+1/2} d\nu\, \ln W_{y,e}(\nu) }\, .
\end{equation}
Hence, the lower bound on the min-entropy arising only from the quantum process is
\begin{equation}
    H_{\rm min} (X \vert E) = -\log_2 \dfrac{\Delta x}{\sqrt{2\pi}\sigma_{x_n}^2} \approx \begin{cases} 10.08~{\rm bit} \text{ for Alice} \\ 9.85~{\rm bit}\text{ for Bob} \end{cases}\, .
\end{equation}
Once $H_{\rm min} (X\vert E)$ has been obtained, the closeness of the hashed homodyne measurements from the ideal uniform distribution is computed by applying the 2-Universal leftover hashing lemma, such that
\begin{equation}
    \varepsilon = 2^{-\frac{1}{2}\left( kH_{\rm min}(X \vert E) - l \right)}\, .
\end{equation}
With the Toeplitz matrix-vector modulo-2 multiplication method that we implemented, the parameters $k$ and $l$ are $32$ and $1$.
The value of $l$ follows from the fact that, although four bits were produced in each round, only one was retained for randomness generation while the remaining three were discarded.
As the extractor is of the strong type, $l$ reflects only the bits actually used.
Under these conditions, the resulting per‑round distance parameter was approximately $10^{-49}$ and $10^{-48}$ for Alice and Bob's QRNG.
Finally, given that the Toeplitz matrix is kept fixed, the distance from the ideal uniform distribution grows linearly with the number of uses of the extractor, reaching a value of $10^{-37}$ and $10^{-36}$ respectively.
Indeed, in our case the experiment ran for approximately \SI{1650}{\second} at a rate of \SI{125}{\mega b/\second}.
\begin{figure}[!ht]
    \centering 
    \includegraphics{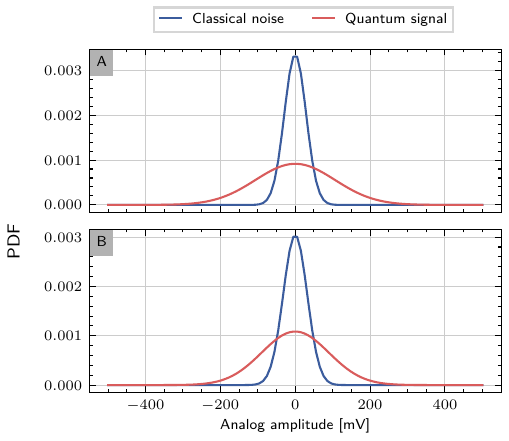}
    \caption{Distribution of the homodyne measured quantum vacuum fluctuations (red curves), (a) Alice and (b) Bob, after the background noise (blue curves) and temporal correlations due to electronics are taken into account.}
    \label{fig:qrng_impz}
\end{figure}
To ensure that the QRNGs were not affected by hardware and digital processing implementation issues, we examined approximately \SI{68}{\giga B} of random bits per device.
To achieve this, we split the \SI{68}{\giga B} of data into \SI{125}{\mega B} files, totaling $511$.
For each file, we run the standard NIST randomness test suite (comprising of $15$ tests) in its default configuration, i.e., the \SI{125}{\mega B} data was divided into $1000$ streams of \SI{1}{\mega ib} each.
For every file, a specific NIST test was considered passed if the p-values of all streams met both the proportion and uniformity criteria.
We considered the original \SI{68}{\giga B} of data to have passed a given NIST test with \SI{99}{\percent} confidence whenever at least $500$ NIST reports indicated that the test was successful.
For both QRNGs, all tests exceeded the threshold, meaning they can be considered passed. Therefore, we can say with a \SI{99}{\percent} confidence interval that the random numbers generated by our devices show no technical issues from hardware or digital processing.

To estimate the window length of the QRNGs, we experimentally estimated the normalized autocorrelation function (ACF) of each device.
By analyzing the latter, we determined the time-lag beyond which the the correlation between measurements becomes negligible, implying that the influence of past homodyne measurements after such a delay are likewise negligible.
This threshold was determined through the normalized cumulative ACF squared function
\begin{equation}
    C_k = \sum_{j=0}^k \rho_j^2 \bigg{/} \sum_{j=0}^N \rho_j^2\, ,
\end{equation}
where $N$ denotes the length of the ACF.
The time‑lag $k_t$ at which $C_k$ reached \SI{99}{\percent} was taken as the threshold.
By doing so, we estimated a time-correlation window of $17$ samples for Bob and $37$ for Alice, corresponding to a time delay of \SI{680}{\pico\second} and \SI{1.48}{\nano\second} given the \SI{25}{\giga S/\second} sampling-rate of the oscilloscope used to acquire the data.

\subsection{System latencies}

A thorough analysis was performed on both the spatial and temporal components of all relevant events of the Bell test.
We began by estimating the distances between the source of entangled photon pairs and the two receiver stations.
To this end, we positioned the source at the spatial origin $(x,y)=(0,0)$ and accounted for all spatial uncertainties in the positions of Alice’s and Bob’s receivers relative to the source.
Direct line-of-sight measurements between the source and the receivers were not feasible due to the building layout. 
Therefore, the distances were calculated based on an architectural blueprint provided by the University of Padova, with a spatial resolution of $\Delta l_0=1$~cm.
Accordingly, the uncertainty assigned to each estimated distance was taken to be at least $N\Delta l_0$, where $N$ denotes the number of segments used in the reconstruction. A conservative estimate was adopted by taking the total uncertainty as $N\Delta l_0$, i.e. by summing the uncertainties of the individual sections linearly rather than in quadrature. 
This corresponds to a worst-case assumption and therefore likely overestimates the true uncertainty.
Although this would have resulted in expected uncertainties below $\sim20$~cm for both spatial coordinates, we opted to conservatively further increase the uncertainty to $50$~cm for both.

\begin{figure*}
    \centering
    \includegraphics{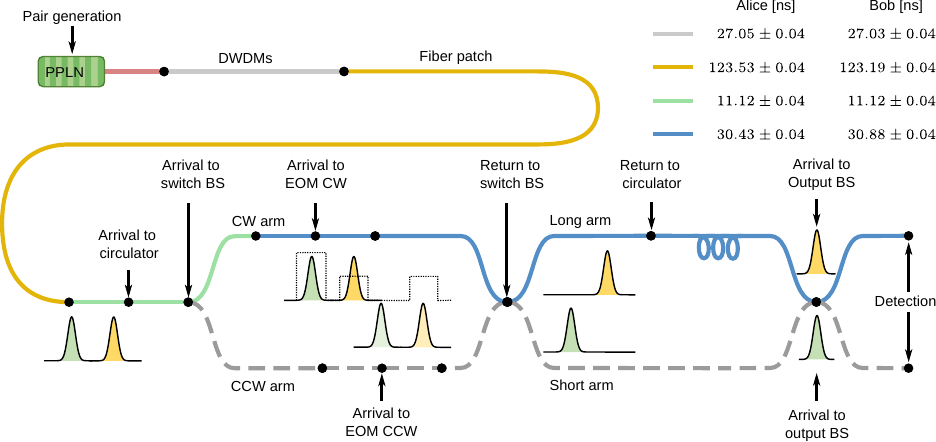}
    \caption{Measured propagation time for the path followed by both signal (Alice) and idler (Bob) photons from the output fiber of the SPDC crystal to the SNSPD input, represented as solid lines. Dashed lines correspond to paths with equal or lower propagation time with respect to the reported values.}
    \label{fig:photon-path-timing}
\end{figure*}

The propagation time through optical fiber was measured through time-resolved measurements using a pair of correlated photons created by a copy of the SPDC source used in the experiment. 
A reference measurement was first performed without the fiber to be measured, where the temporal offset maximizing the coincidence correlation was identified as the reference delay.
The measurement was then repeated after inserting the fiber to be measured into one path, from which the additional propagation delay introduced by the fiber was calculated as the difference between the correlation peak position with and without the fiber.
While this method gives measurements with an uncertainty bounded by the jitter of the SNSPDs (on the order of tens of picoseconds), some portions of the path cannot be directly measured, such as the internal single mode fibers placed from the SNSPD system input to the actual detectors.
These are estimated by measuring the length and assuming a refractive index of $n_{\rm SMF}=1.46$ to estimate the propagation time.
Once again, to go beyond the most pessimistic scenario, we further increased the uncertainty associated with the fiber length to be around $10\%$ its value. 
The corresponding fiber paths measured using this method and their values are displayed on Fig.~\ref{fig:photon-path-timing}.

Lastly, three extra latency components define the time at which we consider the random bit to be generated. First, the propagation times through coaxial cable were simply estimated via time-domain reflectometry, which characterized the connections from the QRNG balanced photodiodes to the FPGA's ADC, and from the FPGA's DAC to the EOM.
Then, the time required for the FPGA to obtain a digital signal from the analog input (ADC to programmable logic), and vice-versa (programmable logic to DAC), combined into a single ADC-DAC loopback parameter $t_{\rm loopback}$, which we measured by developing a custom FPGA circuit to perform the loopback from an ADC input to a DAC output.
Comparing the temporal shift between a $1$~MHz sinusoidal signal after going through the loopback with the signal bypassing it, we obtained $t_{\rm loopback}=103.3\pm0.1$~ns.
Finally, the processing time for the FPGA to perform the Toeplitz extraction, which was implemented to take a total time of $t_{\rm FPGA}=10$~ns, corresponding to a clock period at $500$~MHz and one at $125$~MHz.

Taking all propagation times and electrical latencies into account, each of the three relevant events for both parties can be evaluated.
We define $t^i_S$ as the photon propagation time from pair generation to arrival at the EOM in the Sagnac switch. 
Then, the detection time $t^i_D$ is obtained from the photon propagation time from the EOM to the single photon detectors. 
Commercial SNSPD systems typically include a single-mode fiber connecting the user input connector to the nanowire itself; the lengths of these fibers were provided by the manufacturer for each system used. 
Finally, the time of the measurement setting choice is determined by electrical and processing latencies, as well as the temporal correlations within the QRNG system. 
We define the range of this event from photodiode detection (including temporal correlations) until signal arrival at the FPGA connector, thereby accounting for any electronic noise along the path that could theoretically influence the random bit generation.
All determined events are summarized in Table~\ref{tab:belltest-events}.

\begin{table*}[t]
    \centering
    \begin{tabular}{l c c}
        \textbf{Event} & \textbf{Alice ($i=A$)} & \textbf{Bob ($i=B$)} \\
        \toprule
        $t_C^i$ & ($-12.9\pm 0.5$~m, $23.6\pm 0.5$~m, $57\pm2$~ns) & ($25.1\pm 0.5$~m, $-8.4\pm 0.5$~m, $57\pm 2$~ns) \\
        $t_S^i$ & ($-12.9\pm 0.5$~m, $23.4\pm 0.5$~m, $172\pm 1$~ns) & ($25.3\pm 0.5$~m, $-8.3\pm 0.5$~m, $172\pm1$~ns) \\
        $t_D^i$ & ($-12.8\pm 0.5$~m, $23.8\pm 0.5$~m, $202\pm 2$~ns) & ($25.0\pm 0.5$~m, $-7.4\pm 0.5$~m, $204\pm 2$~ns) \\
        \bottomrule
    \end{tabular}
    \caption{Relevant events of both parties performing a Bell test. Both spatial and temporal coordinates are relative to the position of the source and the creation time of an entangled pair.}
    \label{tab:belltest-events}
\end{table*}